\documentclass{aastex701}

\usepackage{amssymb}
\usepackage{tikz}
\usetikzlibrary{arrows.meta,positioning}
\usepackage{graphicx} 
\usepackage{float}
\usepackage{CJKutf8} 
\usepackage{amsmath,amssymb}   
\usepackage[shortlabels]{enumitem}
\usepackage{xcolor} 
\DeclareRobustCommand{\corrauthorstar}{\texorpdfstring{$^\star$}{*}}

\begin{document}
\begin{CJK*}{UTF8}{gbsn} 

\title{
The FAST-SETI Milky Way Globular Cluster Survey I: A Pilot Multibeam On-the-Fly Search of Five Globular Clusters at L-Band}

\correspondingauthor{Tong-Jie Zhang, Zhen-Zhao Tao}

\author[orcid=0000-0002-8719-3137,sname='Huang']{Bo-Lun Huang (黄博伦)}
\affiliation{Institute for Frontiers in Astronomy and Astrophysics, Beijing Normal University, Beijing 102206, China}
\affiliation{School of Physics and Astronomy, Beijing Normal University, Beijing 100875, China}
\email{Bolunh@hotmail.com}

\author[orcid=0000-0002-4683-5500,sname='Tao']{Zhen-Zhao Tao (陶振钊)\corrauthorstar}
\affiliation{College of Computer and Information Engineering, Dezhou University, Dezhou 253023, China}
\email{tzzzxc@163.com}

\author[orcid=0000-0002-3363-9965,sname='Zhang']{Tong-Jie Zhang (张同杰)\corrauthorstar}
\affiliation{Institute for Frontiers in Astronomy and Astrophysics, Beijing Normal University, Beijing 102206, China}
\affiliation{School of Physics and Astronomy, Beijing Normal University, Beijing 100875, China}
\email{tjzhang@bnu.edu.cn} 

\author[orcid=0000-0002-8604-106X,sname='Gajjar']{Vishal Gajjar}
\affiliation{University of California, Berkeley, 501 Campbell Hall 3411, Berkeley, CA 94720, USA}
\affiliation{SETI Institute, 339 Bernardo Ave, Suite 200, Mountain View, CA 94043, USA}

\email{vishalg@berkeley.edu}

\footnotetext[1]{Corresponding authors: Tong-Jie Zhang (\texttt{tjzhang@bnu.edu.cn}) and Zhen-Zhao Tao (\texttt{taozhenzhao@dzu.edu.cn})}

\begin{abstract}
We report a narrowband technosignature search toward five Milky Way globular clusters---NGC~6171, NGC~6218, NGC~6254, NGC~6838, and IC~1276---using FAST’s 19-beam L-band receiver (1.05--1.45\,GHz). We adapted MultiBeam Point-source Scanning (MBPS) to extended targets by gating detections to generalized on-target windows (gOTWs), i.e., the time intervals when a beam’s main lobe intersects a buffered cluster mask, and by enforcing the deterministic multibeam illumination sequence as a geometry test. Spectra at $\Delta\nu \approx 7.5$\,Hz resolution and $\Delta t \approx 10$\,s cadence were searched with \texttt{turboSETI} over $|\dot{\nu}|\le 4~\mathrm{Hz\,s^{-1}}$ at $\mathrm{S/N}\ge 10$. From $2.75\times10^{5}$ raw hits across both linear polarizations, none survived the gOTW gating, array-wide simultaneity veto, in-stripe ordering, and single-drift coherence checks, yielding a robust null result. With $\mathrm{SEFD}\approx 1.5$\,Jy and an effective $\sim60$\,s per illuminated crossing, our per-crossing \textcolor{darkgray}{flux density} threshold is $S_{\min}\approx0.50$\,Jy, corresponding to \textcolor{darkgray}{$\mathrm{EIRP}_{\min}\approx (0.72\text{--}1.8)\times10^{16}$\,W} for cluster distances of 4--6.5\,kpc; when multiple illuminated crossings occur, non-coherent stacking improves sensitivity by up to $\sqrt{N}$. To our knowledge, this is the first FAST technosignature survey dedicated to globular clusters and the first to use MBPS as the primary observing strategy. These limits disfavour bright, persistent, isotropic L-band beacons above the stated thresholds during our epochs and establish a scalable blueprint---geometry-aware gating and verification---for multi-epoch MBPS campaigns that expand signal morphologies and combine passes to deepen constraints on transmitters in dense stellar systems.
\end{abstract}


\section{Introduction}

The search for technosignatures—observable indicators of technology developed by extraterrestrial intelligence—has long emphasized radio transmissions, particularly ultra–narrowband features (Hz–level) that are widely treated as hallmarks of engineered emitters rather than natural astrophysical processes. This logic underpins many modern radio SETI programs, which prioritize searches for narrowband carriers across GHz bands \citep[e.g.,][]{Siemion2013Kepler,Enriquez2017BL692,Price2020BL1327}.

Despite decades of effort, no confirmed extraterrestrial signal has been identified. A central challenge is the ubiquity of terrestrial radio–frequency interference (RFI), which can closely mimic sky–localized technosignatures and inflate candidate lists by orders of magnitude before vetting \citep[e.g.,][]{Enriquez2017BL692,Price2020BL1327}. These experiences motivate observing strategies and pipelines that discriminate RFI in situ and at scale.

To that end, a MultiBeam Point-Source Scanning (MBPS) strategy has been proposed and tested with the Five-hundred-meter Aperture Spherical radio Telescope (FAST). MBPS exploits a multibeam receiver and controlled slews so that a sky-localized signal appears only in the beam presently aligned with the source, whereas broad RFI typically enters multiple beams simultaneously. In a FAST trial, MBPS logic rejected all initial narrowband candidates as RFI, illustrating the approach’s power to suppress persistent false positives \citep{Huang2023MBPS}.

FAST is exceptionally well suited to such strategies. Its 19-beam L-band receiver (1.05–1.45\,GHz) and low system temperature deliver outstanding sensitivity for technosignature searches \citep{Jiang2020FAST,Nan2011,Li2016}. The multibeam architecture is integral to MBPS and enables simultaneous on-source/off-source comparisons within a single pointing pattern.

Beyond nearby stars, there is growing interest in extending searches to diverse environments. Globular clusters (GCs) are ancient, gravitationally bound stellar systems with $10^{5}$–$10^{6}$ members; their high stellar densities could shorten communication and travel times within a putative technological network, yet crowding and typically low metallicity may inhibit planet formation and long-term orbital stability \citep{FischerValenti2005,BeerKingPringle2004,Sigurdsson2003,DiStefano2016}. From an observational perspective, GC demographics and structural parameters are well compiled in the Harris catalog and updates, which inform experiment design \citep{harris1996}.

This work presents the first FAST SETI survey dedicated to GCs and the first to employ MBPS as the primary observing strategy. We target NGC~6171 (M107), NGC~6218 (M12), NGC~6254 (M10), NGC~6838 (M71), and IC~1276 (Pal~7). Our pilot survey demonstrates MBPS in extended, dense stellar fields and places sensitive upper limits on powerful radio transmitters within these systems. Section~\ref{sec:methods} summarizes observations and configuration; Section~\ref{sec:data} details search methods, RFI assessment, and sensitivity; Section~\ref{sec:discussion} discusses implications and prospects for future GC technosignature searches.

\section{Observations}\label{sec:methods}

\subsection{Target Selection}\label{sec:targets}

\textit{Rationale.} Globular clusters (GCs) are ancient, dense stellar systems that offer both astrophysical motivation and practical advantages for technosignature searches. Theoretical work has argued that, despite high stellar encounter rates in cluster cores, there exist substantial “sweet-spot’’ regions in many GCs where habitable-zone orbits can remain dynamically stable over gigayear timescales; the short interstellar separations in such regions would also greatly facilitate interstellar communication and travel for any advanced civilization \citep{DiStefano2016}. At the same time, the well-established planet--metallicity correlation suggests that stars drawn from the metal-richer tail of the GC distribution should be more likely to host giant planets (and perhaps complex planetary architectures) than their more metal-poor counterparts \citep{FischerValenti2005}. Empirically, planets can exist in GCs: the hierarchical triple in M4 (PSR~B1620$-$26 with a planetary-mass companion) provides a long-discussed proof-of-concept that planetary bodies can survive, or be (re)assembled, in low-metallicity cluster environments \citep{Sigurdsson2003,BeerKingPringle2004}. Together, these considerations motivate a focused SETI experiment on relatively metal-rich Milky Way GCs.

\textit{Selection criteria.} We therefore constructed a five-object sample --- NGC~6171 (M107), NGC~6218 (M12), NGC~6254 (M10), NGC~6838 (M71), and IC~1276 (Pal~7) --- optimized along four axes:

\begin{enumerate}
\item \emph{Metallicity (planet formation prior).} We prioritized clusters toward the metal-rich end of the Galactic GC distribution, where the prior probability for giant-planet occurrence is higher \citep{FischerValenti2005}. In particular, M71 and Pal~7 are among the more metal-rich Milky Way GCs, while M10/M12/M107 are of intermediate metallicity. All metallicities quoted in this paper are drawn uniformly from the Harris catalog (1996; 2010 edition) to ensure internal consistency \citep{harris1996}.

\item \emph{Proximity (EIRP sensitivity).} For a flux-limited narrowband search, the minimum detectable equivalent isotropic radiated power scales as $d^2$. We therefore favored nearby clusters (heliocentric distances $\sim$4--6.5~kpc), improving our EIRP limits for any continuous narrowband transmitters \citep{harris1996}. 

\item \emph{Stellar population size (trial factor).} Integrated absolute magnitudes ($M_V$) from the Harris catalog were used as proxies for total stellar population. Brighter clusters (e.g., M10, M12, M107 with $M_V\sim-7$) host larger numbers of stars, increasing the chance of an $n{=}1$ technosignature source within the beam at any instant \citep{harris1996}.

\item \emph{Observational accessibility with FAST.} FAST’s L-band 19-beam system delivers its highest gain for elevations near zenith and provides high stability across 1.05--1.45~GHz \citep{Jiang2020}. Its geographic latitude (25\fdg8~N) and sky-coverage constraints (zenith angle $\lesssim40^\circ$, roughly $-14^\circ\!\lesssim\delta\!\lesssim\!+65^\circ$) comfortably include our targets \citep{Li2016,Nan2011}. We scheduled each GC near meridian transit to minimize air mass and leverage the instrument’s best sensitivity.
\end{enumerate}

\textit{Scientific expectations.} From an SETI perspective, this sample spans a scientifically useful range of cluster environments while keeping the priors favorable. The inclusion of comparatively metal-rich disk/bulge clusters (M71, Pal~7) is directly aligned with planet-occurrence evidence \citep{FischerValenti2005}, and the presence of the PSR~B1620$-$26 planetary system in M4 demonstrates that planetary companions (and complex dynamical assembly channels) are realizable in GCs \citep{Sigurdsson2003,BeerKingPringle2004}. Theoretical work indicates that, even in dynamically active systems, large volumes outside the densest cores can host long-lived habitable zones \citep{DiStefano2016}. Observationally, all five targets are close enough that FAST’s sensitivity yields stringent EIRP limits for continuous, narrowband carriers (see \S\ref{sec:sensitivity}), and their declinations place them in FAST’s high-gain sky. For reproducibility and uniformity, the equatorial coordinates, distances, metallicities, and $M_V$ values used here are taken from the Harris (2010 edition) catalog \citep{harris1996}; the specific parameter values and the observation windows are listed in Table~\ref{tab:gc_params} and Table~\ref{tab:obs_log}.

\begin{deluxetable*}{lccccccc}
\tablecaption{Target Globular Clusters and Key Properties (from \citealp{harris1996})\label{tab:gc_params}}
\tablehead{
\colhead{Cluster} & \colhead{Alt. Name} & \colhead{RA (J2000)} & \colhead{Dec (J2000)} & \colhead{Distance (kpc)} & \colhead{[Fe/H]} & \colhead{$M_V$ (mag)} & \colhead{Angular size(arcmin)}
}
\startdata
NGC\,6171 & M107       & 16$^{\rm h}$32$^{\rm m}$31$^{\rm s}$ & $-13^\circ03'14''$ & 6.4 & $-1.02$ & $-7.12$ & $10.0$\\
NGC\,6218 & M12        & 16$^{\rm h}$47$^{\rm m}$14$^{\rm s}$ & $-01^\circ56'55''$ & 4.8 & $-1.37$ & $-7.31$ & $14.5$\\
NGC\,6254 & M10        & 16$^{\rm h}$57$^{\rm m}$09$^{\rm s}$ & $-04^\circ06'01''$ & 4.4 & $-1.56$ & $-7.48$ & $15.1$\\
NGC\,6838 & M71        & 19$^{\rm h}$53$^{\rm m}$46$^{\rm s}$ & $+18^\circ46'45''$ & 4.0 & $-0.78$ & $-5.61$ & $7.2$\\
IC\,1276  & Palomar\,7 & 18$^{\rm h}$10$^{\rm m}$44$^{\rm s}$ & $-07^\circ12'27''$ & 5.4 & $-0.75$ & $-6.67$ & $19.2$\\
\enddata
\tablecomments{Harris (1996; 2010 edition) values. Distances are heliocentric; metallicities in the Zinn--West scale. Angular sizes are from \cite{1989Sci}}
\end{deluxetable*}.

\begin{deluxetable*}{lccccccc}
\tablecaption{FAST MBOTF Observing Log and Scan Geometry\label{tab:obs_log}}
\tablehead{
\colhead{Cluster} & \colhead{Date (UTC)} & \colhead{Start} & \colhead{End} & \colhead{Dur. (s)} & \colhead{First RA Sweep} & \colhead{$\Delta$Dec ($'$)} & \colhead{Second RA Sweep}
}
\startdata
NGC\,6838 (M71)   & 2024-07-10 & 00:41:30 & 01:06:22 & 1492 & Westward, 33.2$'$ & 2.49 & Eastward, 33.2$'$ \\
IC\,1276 (Pal\,7) & 2024-07-09 & 23:57:38 & 00:29:30 & 1912 & Westward, 45.2$'$ & 2.49 & Eastward, 45.2$'$ \\
NGC\,6254 (M10)   & 2024-07-09 & 23:16:40 & 23:45:38 & 1738 & Westward, 41.1$'$ & 2.49 & Eastward, 41.1$'$ \\
NGC\,6218 (M12)   & 2024-07-09 & 22:38:08 & 23:06:40 & 1712 & Westward, 40.5$'$ & 2.49 & Eastward, 40.5$'$ \\
NGC\,6171 (M107)  & 2024-07-09 & 22:00:00 & 22:26:08 & 1568 & Westward, 36.0$'$ & 2.49 & Eastward, 36.0$'$ \\
\enddata
\tablecomments{Slew speed $3''$\,s$^{-1}$. The first RA sweep passes through the cluster-center declination; the second RA sweep is offset by $\sim$2.49$'$. For all entries, the first sweep is westward, hence the OTB order is 8$\rightarrow$2$\rightarrow$1$\rightarrow$5$\rightarrow$14.}
\end{deluxetable*}

\begin{figure*}[htbp!]
\centering
\includegraphics[width=0.9\textwidth]{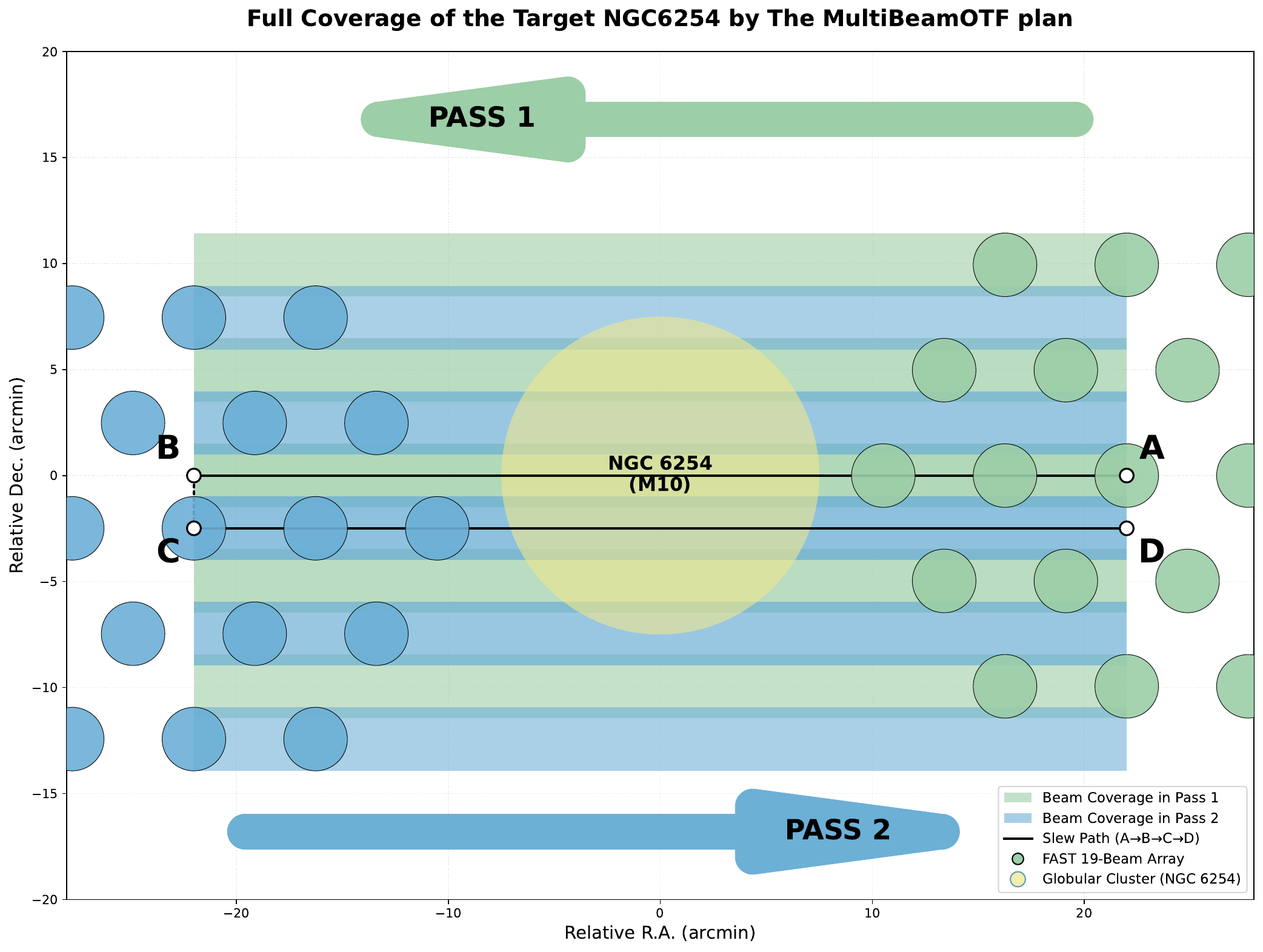}
\caption{Schematic illustration of the OTF raster scan strategy, an adaptation of the MBPS technique for extended targets like the globular cluster NGC 6254 (central yellow circle). The observation follows the precise path A→B→C→D to ensure complete coverage.
The first scan (\textbf{Pass 1}) commences at point \textbf{A}, with the 19-beam array (cyan) slewing in the negative right ascension direction to point \textbf{B}. A single, precise step in declination ($\Delta\text{Dec} \simeq 2.49'$) is then executed to move the array to point \textbf{C}. From here, the second scan (\textbf{Pass 2}) begins, slewing in the reverse right ascension direction (coral) and concluding the observation at point \textbf{D}.
The shaded cyan and coral regions illustrate the sky coverage traced by each of the 19 beams. This two-pass methodology provides complete and overlapping coverage of the target's full angular extent. Crucially, this strategy preserves the core principle of MBPS: a genuine extraterrestrial signal must appear in the strict temporal and spatial sequence of on-target beam crossings within each pass, a pattern that terrestrial RFI is highly unlikely to replicate.}
\label{fig:scan_strategy}
\end{figure*}

\begin{figure*}[htbp!]
\centering
\includegraphics[width=0.7\textwidth]{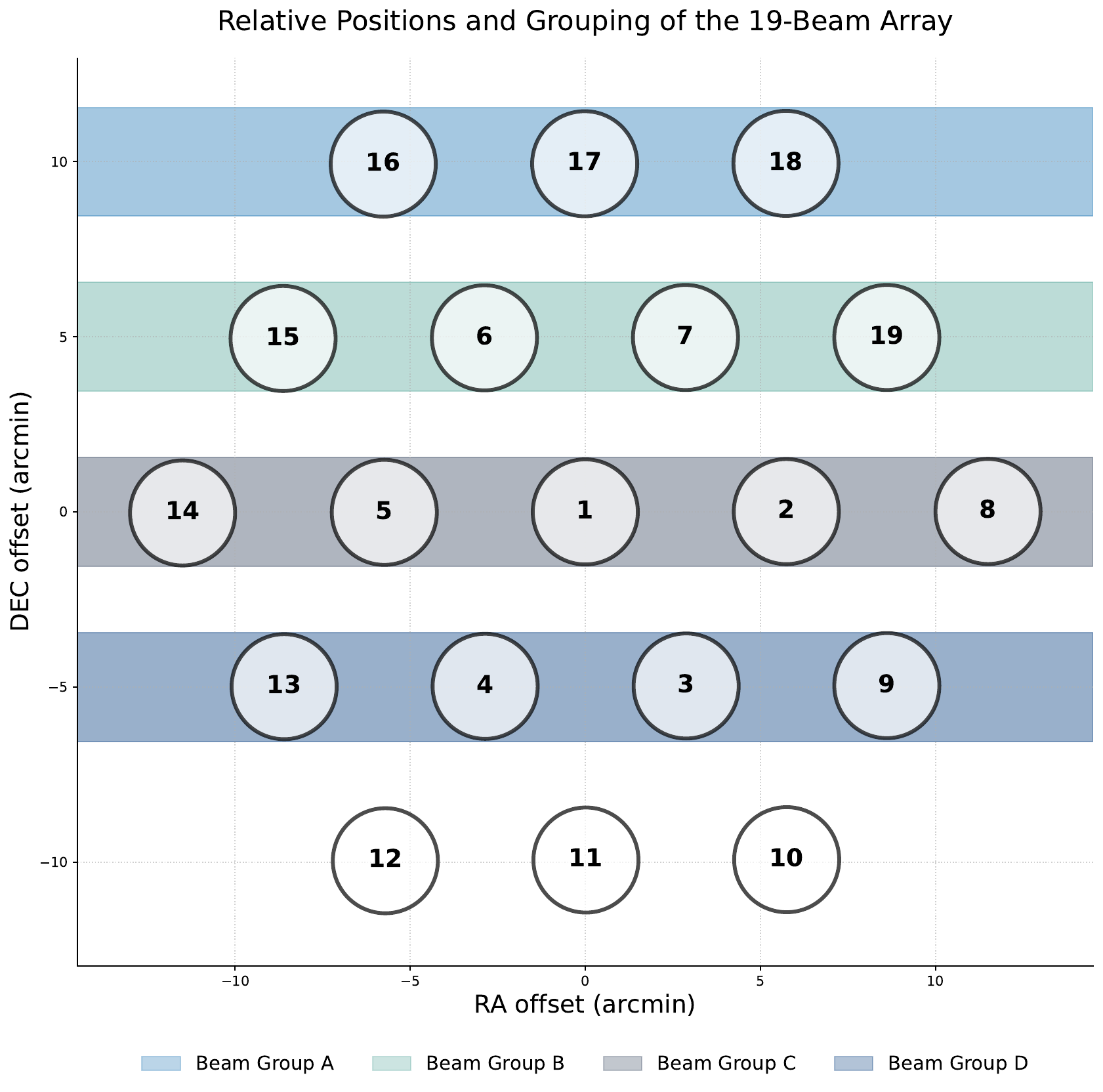}
\caption{The on-sky layout of the 19-beam L-band receiver of the Five-hundred-meter Aperture Spherical radio Telescope (FAST). For the observation presented in this work, the beams were organized into four distinct Beam Groups (BGs), labeled A through D. This grouping corresponds to the sequence in which different sets of beams scanned across the target globular cluster. Each group is highlighted by a unique, semi-transparent shaded region, with the color coding specified in the legend.}
\label{fig:beamgroups}
\end{figure*}

\subsection{FAST Instrument, MBOTF Scanning, and Geometry}\label{sec:mbotf-geometry}

Our observations used the FAST 19-beam L-band receiver (1.05–1.45\,GHz) with the SETI backend, configured at spectral resolution $\Delta\nu\simeq7.45$\,Hz and integration time $\Delta t\simeq10.20$\,s; both linear polarizations were recorded. Because the half-power beamwidth (HPBW) of a single FAST beam at 1.3\,GHz ($\sim$3′) is smaller than the angular extents of our targets, we adopted a multibeam on-the-fly (MBOTF) “snake” raster along right ascension (RA) at a constant slew speed of $3''\,{\rm s}^{-1}$. The receiver rotation was set to $0^\circ$, which aligns five beams nearly colinear along constant declination; these five beams form the familiar on–target sequence used in point-source MBPS contexts (beams \{8,\,2,\,1,\,5,\,14\} at $0^\circ$ rotation).

Each observation consists of two RA sweeps (Pass~1 and Pass~2). Pass~1 tracks the cluster-center declination. The telescope then steps in declination by $\Delta{\rm Dec}\simeq2.49'$ (approximately the adjacent-beam spacing) and executes Pass~2 in the opposite RA direction. The union of the two passes provides continuous, overlapping coverage across the globular-cluster field: some sky positions are illuminated in both passes, while others are visited in only one, depending on their declination within the raster.

Within a single sweep, the cluster center crosses the on–target five-beam line in a deterministic order that depends on scan direction (for a westward sweep, $8\rightarrow2\rightarrow1\rightarrow5\rightarrow14$; for an eastward sweep the order reverses). The effective dwell time for a beam-center crossing is
\[
t_{\rm OTW}\approx \frac{{\rm HPBW}}{\dot{\theta}}\;\approx\;\frac{3'}{0.05'\,{\rm s}^{-1}}\;\approx\;60~{\rm s},
\]
and the separation between adjacent on–target beam centers ($\sim$5.74′) implies $\sim$115\,s between consecutive crossing midpoints at the commanded scan rate. These deterministic spacings (in angle and in time), together with the two-pass complementary coverage, are the geometric constraints that underpin our later multibeam verification (§\ref{sec:mbps-verify}).

For subsequent analysis across the full 19-beam footprint, the array tracks four adjacent declination stripes during the raster. We therefore refer to four \emph{beam groups} (BGs), A–D, each denoting one stripe traced by a subset of beams in a pass. The central stripe (BG–C) contains the five-beam line that passes the cluster core; the adjacent stripes (BG–A/B/D) provide complementary illumination above and below BG–C. This stripe-based view is purely geometric and does not change the instrument configuration; it serves to organize timing/order expectations and coverage accounting in the cross–verification stage (§\ref{sec:mbps-verify}).

\section{Data Analysis}\label{sec:data}

\subsection{Primary Search Configuration and Gating}\label{sec:search-setup}

The primary narrowband search is performed across all four beam groups (BGs) rather than a limited on–target subset, because a globular cluster is an extended source.  Any beam whose main lobe overlaps the cluster region during the scan is relevant for detection and subsequent geometric verification.

During the MBOTF raster, the FAST 19-beam receiver at a rotation angle of $0^\circ$ traces four adjacent declination stripes.  These are denoted BG–A through BG–D, each corresponding to one stripe.  BG–C passes through the cluster core and corresponds to the five beams traditionally referred to as on–target beams in point-source MBPS work; BG–A, B, and D cover the adjacent stripes.  All subsequent geometric tests are applied per BG and per scan pass.

Two right-ascension passes are executed, separated by the nominal MBOTF declination step.  Together, Pass 1 and Pass 2 provide continuous coverage of the buffered cluster mask; some regions are illuminated in both passes, while others only once.  Beams outside their illuminated windows at a given time serve as reference channels for array-wide vetoes.

For each cluster, a sky mask is generated from its optical boundary and expanded by the L-band half-power beamwidth (HPBW) to define the buffered region.  For every beam $b$, the time intervals when its boresight intersects this region constitute its generalized on–target windows (gOTWs).  Searches are restricted to these intervals; data outside gOTWs are excluded from detection but retained for simultaneity tests.

Spectra are analyzed at $\sim$7.5 Hz frequency resolution and $\sim$10 s integration.  Linear polarizations (XX and YY) are processed independently over a drift-rate range $|{\dot \nu}|\le4$ Hz s$^{-1}$, and events with ${\rm S/N}\ge10$ are recorded.  The resulting per-beam hit lists contain the BG label, pass index, gOTW identifier, boresight offset from the mask, and boundary flags marking hits within a few seconds of gOTW edges.  Geometry-based validation is applied afterward (§\ref{sec:mbps-verify}).  A conservative HPBW corresponding to the widest beam in 1.05–1.45 GHz is used to ensure uniform gating across the band.


\subsection{MBPS Cross-Verification Using Beam Groups}\label{sec:mbps-verify}

Geometric verification follows the deterministic MBPS pattern within each beam group (BG) and across both complementary passes to identify signals consistent with sky-fixed sources. We apply the following tests, referred to below as (C1)–(C5).

\textbf{(C1) gOTW gating (per BG, per pass).} Candidate formation is restricted to intervals when a beam is illuminated, i.e., within its generalized on–target window (gOTW) defined by the intersection of the beam track with the buffered cluster mask. Hits outside gOTWs are excluded from geometry but are retained as reference channels.

\textbf{(C2) Array-wide simultaneity veto.} At any time, beams outside their gOTWs on any BG and pass serve as reference beams. Signals detected simultaneously across many reference beams (or across BGs irrespective of gOTW gating) are classified as terrestrial interference. This veto removes array-illuminating features including zero-drift carriers and broad spectral structures.

\textbf{(C3) In-stripe ordering and timing (per BG, per pass).} Within a BG stripe, the beams sweep a sky position in a fixed spatial and temporal order determined by the scan rate and inter-beam spacing. A genuine emitter inside the buffered mask should recur across the illuminated beams of that BG with the expected sequence and $\sim115$\,s separations for that pass and scan direction. Candidates are required to satisfy this schedule in at least $N_{\rm min}$ crossings when geometry predicts that many; violations of order or timing lead to rejection as geometrically inconsistent.

\textbf{(C4) Drift coherence and beam-response plausibility.} Surviving candidates must admit a single drift-rate solution $\dot{\nu}$ across their crossings, with residuals consistent with the search resolution, and exhibit amplitudes compatible with the main-lobe response within gOTWs (peaking nearer boresight and tapering toward edges). Deviations attributable to partial illumination near gOTW boundaries are tolerated.

\textbf{(C5) Cross-pass consistency under complementary coverage.} Because Pass~1 and Pass~2 illuminate complementary declination stripes, absence of a recurrence in the other pass is not penalized if the position is not illuminated there. Where a position is illuminated in both passes, the candidate must show the same $\dot{\nu}$ and satisfy the in-stripe schedule on each pass.

Polarizations are analyzed independently; coincidence between XX and YY is not required but is recorded for diagnostics. For candidates that satisfy (C1)–(C5), we apply non-coherent stacking of detections from multiple illuminated crossings (and passes when applicable) using inverse-variance weights; the expected sensitivity improvement scales as $\sqrt{N_{\rm cross}}$.

\begin{figure}[t]
\centering
\includegraphics[width=0.95\textwidth]{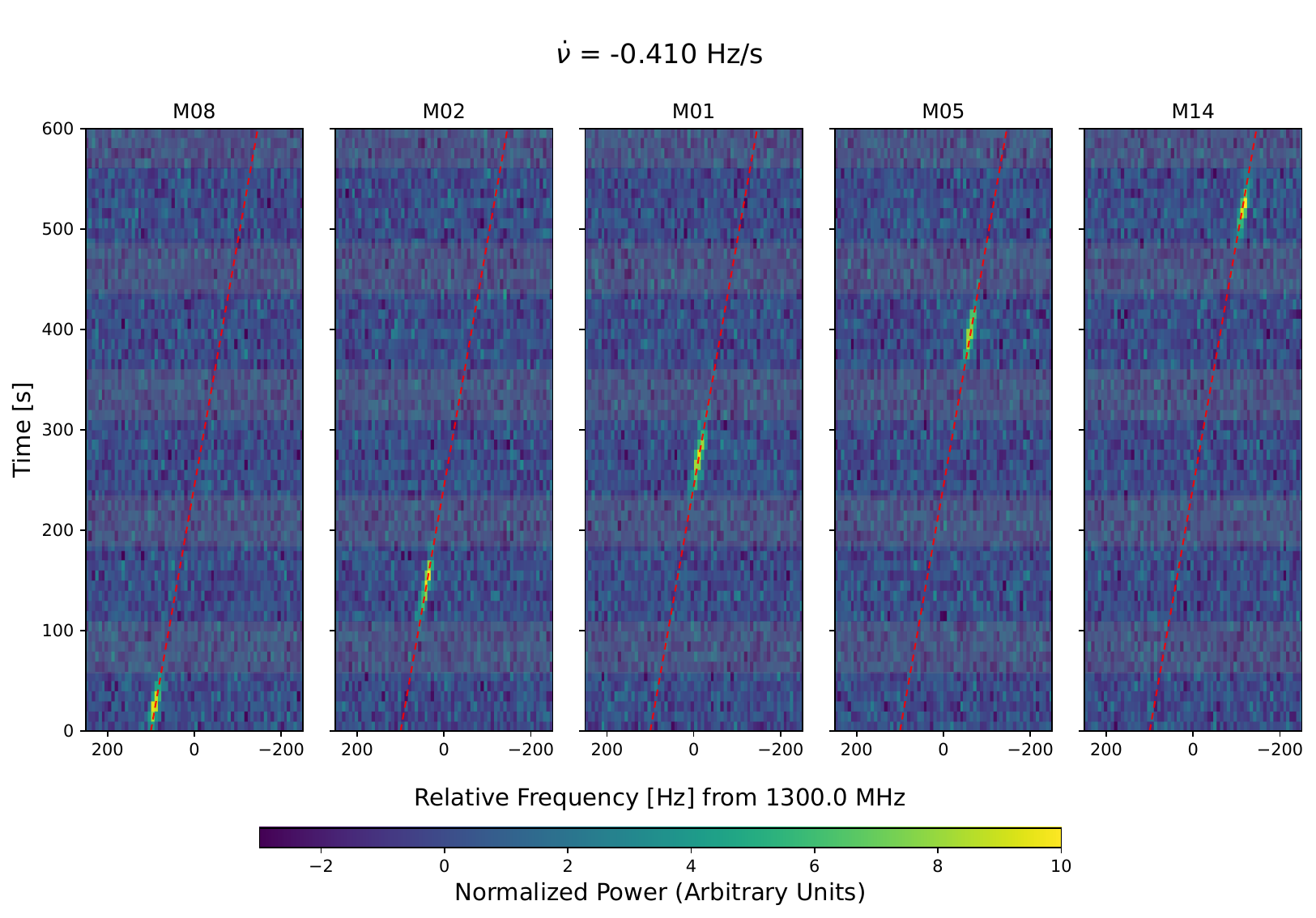}
\caption{
\textbf{Simulated appearance of a geometrically consistent narrowband signal under MBPS scanning.}
Each subpanel (M08, M02, M01, M05, M14) represents one beam of a single beam group (BG) during a raster scan, showing normalized power versus time and relative frequency from 1300\,MHz. 
The injected tone drifts at $\dot \nu=-0.41$\,Hz\,s$^{-1}$ and is visible sequentially as the main lobe of each beam sweeps over the source location. 
The recurrence pattern, $\sim115$\,s timing separations, and consistent drift rate satisfy the MBPS geometric criteria (C1–C4): the signal appears only inside gOTWs, does not illuminate all beams simultaneously, recurs with the correct in-stripe order, and maintains a single drift-rate solution with amplitudes following the beam-response envelope. 
This example illustrates what a genuine sky-localized technosignature would look like under the FAST MBPS observation strategy.
}
\label{fig:simulation}
\end{figure}

\subsection{Sensitivity and EIRP}\label{sec:sensitivity}
For a narrowband, unresolved tone detected within a single gOTW, the minimum detectable
flux density follows the radiometer equation,
\begin{equation}
    S_{\min} \;=\; \frac{{\rm SEFD}\,\times\,({\rm S/N})}{\sqrt{N_{\rm pol}\,\Delta\nu\,t_{\rm int}}}\,,
    \label{eq:radiometer}
\end{equation}
with ${\rm SEFD}=1.5$\,Jy, ${\rm S/N}=10$, $N_{\rm pol}=2$, channel width $\Delta\nu \simeq 7.5$\,Hz, and
effective integration per crossing $t_{\rm int}\simeq t_{\rm gOTW}\simeq 60$\,s. This yields
$S_{\min}\simeq 0.50$\,Jy for a single OTB crossing.\footnote{Using $N_{\rm pol}=1$ would give
$S_{\min}\simeq 0.71$\,Jy; throughout we adopt $N_{\rm pol}=2$ and propagate that choice consistently.}

\textcolor{darkgray}{For an unresolved spectral line, we convert this flux density into a total EIRP within a single 7.5\,Hz channel as}
\begin{equation}
    \textcolor{darkgray}{{\rm EIRP}_{\min} \;=\; 4\pi d^{2}\,S_{\min}\,\Delta\nu}\,,
    \label{eq:eirp}
\end{equation}
\textcolor{darkgray}{where $d$ is the cluster distance and $\Delta\nu=7.5$\,Hz is the spectral resolution.} With $S_{\min}\!=\!0.50$\,Jy and $d\!=\!4$--$6.5$\,kpc (Table~\ref{tab:gc_params}),
\textcolor{darkgray}{we obtain ${\rm EIRP}_{\min}\!\approx\!(0.72\text{--}1.8)\times10^{16}$\,W.} These values apply to a single,
on--center OTB crossing in Pass~1.

\paragraph{Stacking across OTB crossings.}
During Pass~1 the five OTB boresight tracks sweep a single declination stripe. 
A sky--fixed emitter located on (or sufficiently close to) this stripe will be crossed sequentially by multiple OTBs; 
in the ideal central geometry this yields \emph{up to} five gOTWs in one pass. 
More generally, the number of usable crossings satisfies $1 \le N_{\rm cross} \le 5$ and depends on the emitter’s 
position relative to the OTB tracks and masking. 
We therefore combine the available OTWs incoherently with inverse--variance weights,
\[
S_{\min,\,{\rm stack}} \;=\; 
\left(\sum_{i=1}^{N_{\rm cross}} \frac{1}{S_{\min,i}^{2}}\right)^{-1/2}
\;\approx\;
\frac{S_{\min}}{\sqrt{\sum_{i=1}^{N_{\rm cross}} w_i}}\,,
\]
where $S_{\min,i}$ includes per--OTW masking and beam--gain factors and $w_i$ denotes the corresponding normalized weights. 
For equal, unmasked, on--center crossings this reduces to $S_{\min}/\sqrt{N_{\rm cross}}$; hence the familiar $\sqrt{5}$ 
gain is an \emph{upper bound}. 
Pass~2, offset by $\Delta{\rm Dec}\!\approx\!2.49'$ and slewing in the opposite RA direction, interleaves with Pass~1 and 
can supply additional OTWs for sources that fell in the Pass~1 gaps; when used for sensitivity, those OTWs are weighted by 
their beam--gain factors (otherwise Pass~2 serves as an RFI cross--check).

\subsection{Choice of Maximum Drift Rate}\label{subsec:drift}

We adopted a maximum drift rate of $\dot \nu_{\max}=\pm4~\mathrm{Hz~s^{-1}}$, equivalent to a normalized ceiling of
\begin{equation}
\eta \equiv \frac{\dot \nu}{\nu} = \frac{4}{1.42\times10^9} \approx 2.8\times10^{-9}\,\mathrm{s^{-1}} = 2.8~\mathrm{nHz}\quad\text{(at 1.42\,GHz)}.
\end{equation}

This choice rests on three considerations:

\begin{enumerate}
\item \emph{Astrophysical plausibility.} Earth’s orbital and rotational accelerations yield drifts $\lesssim0.2~\mathrm{Hz~s^{-1}}$ at L-Band \citep{Siemion2013Kepler}. A transmitter on an Earth-like planet would plausibly add a similar contribution, placing realistic drifts in the $\sim0.3$--$0.5~\mathrm{Hz~s^{-1}}$ range. Our ceiling exceeds this by nearly an order of magnitude, comfortably encompassing Earth-like and moderately more extreme kinematics.

\item \emph{Algorithmic feasibility.} With $\Delta\dot \nu=0.125~\mathrm{Hz~s^{-1}}$, $\pm4~\mathrm{Hz~s^{-1}}$ implies $\sim65$ trials per scan, a computationally efficient load for multi-hour FAST datasets. Extending to the ``200\,nHz'' normalized envelope ($\dot \nu_{\max}\!\sim\!280~\mathrm{Hz~s^{-1}}$ at 1.4\,GHz; \citealt{Sheikh2020ETZ}) would inflate trials by a factor of $\sim70$, substantially increasing runtime and false-positive candidates without clear gains for 60\,s scans.

\item \emph{Comparative precedent.} Our choice aligns with Breakthrough Listen’s $\pm4~\mathrm{Hz~s^{-1}}$ searches of nearby stars \citep{Enriquez2017BL692}, exoplanet hosts \citep{Price2020BL1327}, and the Galactic Center \citep{Gajjar2021GC}, facilitating direct comparability of sensitivity and candidate rates across diverse astrophysical contexts, now extended to GCs.
\end{enumerate}

\section{Results}\label{sec:results}

\subsection{TurboSETI Hit Statistics}\label{sec:hits}
Using the configuration in \S\ref{sec:data} (threshold ${\rm S/N}=10$, $\Delta\nu\simeq7.5$\,Hz, $|\dot{\nu}|\le 4$\,Hz\,s$^{-1}$), 
\texttt{turboSETI} reported a total of $2.75\times10^{5}$ raw hits across the five targets and both polarizations 
(XX: 143{,}890; YY: 130{,}875; Table~\ref{tab:hits}). XX and YY totals agree within $\sim10\%$, indicating that most 
interferers are not strongly polarized. Frequency histograms show pronounced concentrations within known aviation 
(SSR 1030--1140\,MHz) and GNSS allocations (e.g., 1176/1207/1227/1268/1381\,MHz), consistent with the site environment 
described in \S\ref{sec:rfi}.  

\begin{deluxetable}{lcccccc}[H]
\tablecaption{Summary of hits from 19 beams per polarization and Target.\label{tab:hits}}
\tablehead{
\colhead{ID.} & \colhead{NGC6838} & \colhead{IC1276} & \colhead{NGC6171} & \colhead{NGC6218} & \colhead{NGC6254} & \colhead{Total}
}
\startdata
XX & 17238 & 31186 & 30841 & 38426 & 26199 & 143890 \\
YY & 16993 & 32026 & 31332 & 26291 & 24233 & 130875 \\
\enddata
\tablecomments{Raw hit counts from \texttt{turboSETI}. Totals include both polarizations: 274765 hits in total across all five globular clusters.}
\end{deluxetable}

\begin{figure*}[htbp!]
\centering
\includegraphics[width=0.9\textwidth]{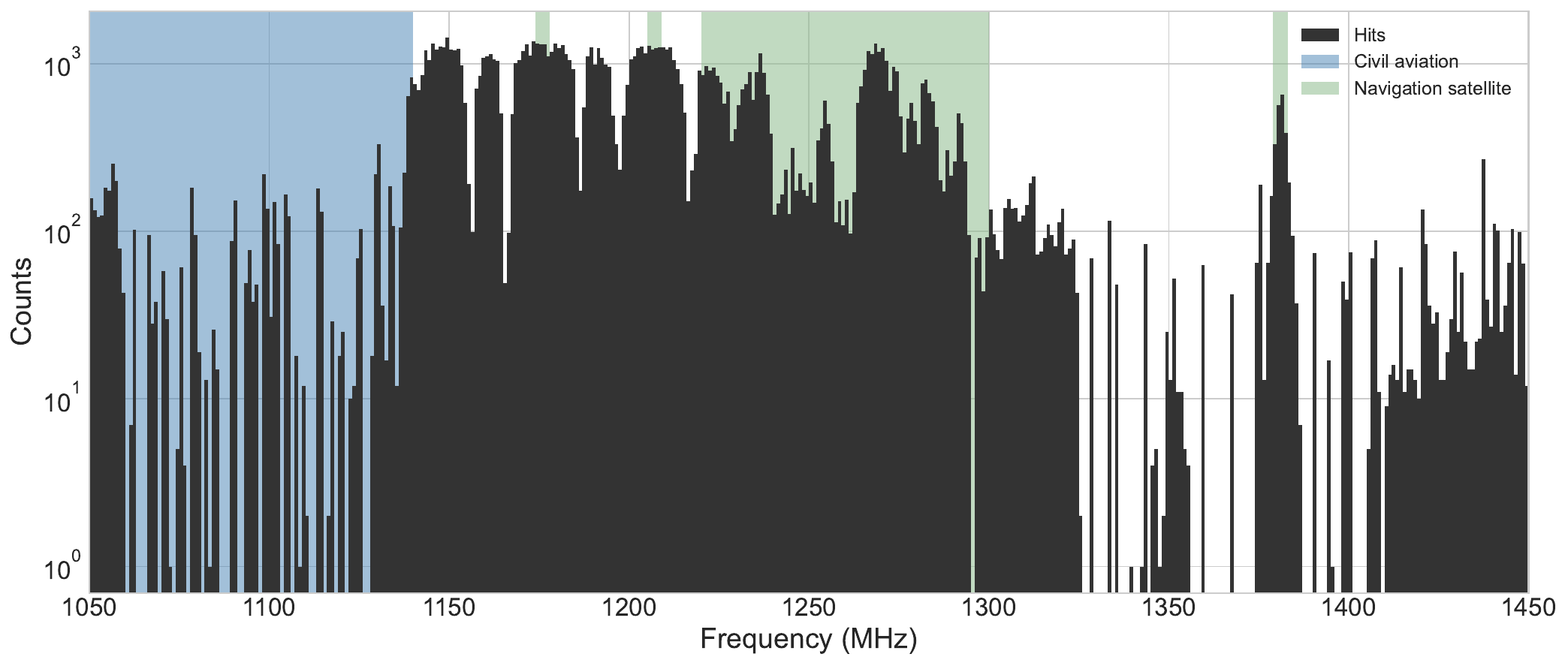}
\caption{Distribution of signal hits detected across the five GCs. The hit density is extremely high in two primary regions. A saturated block of hits from 1220--1300\,MHz corresponds to the navigation satellite band (green), rendering it unusable. A dense, but resolved, cluster of hits aligns with the civil aviation allocation (blue, 1030--1140\,MHz). Notably, a strong, isolated hit peak at 1381.05\,MHz is also explained by a navigation satellite downlink (GPS L3),}
\label{fig:rfi}
\end{figure*}
\subsection{MBPS Filtering Outcomes and Null Result}\label{sec:mbps-null}

We apply the verification framework of §\ref{sec:mbps-verify} to the per-beam hit lists from §\ref{sec:search-setup}. In the \emph{operational} order of the pipeline, the filtering proceeds by enforcing (C1)–(C5): (C1) gOTW gating per BG and pass (restricting candidate formation to illuminated intervals), then (C2) the array-wide simultaneity veto using reference beams (outside their gOTWs), followed by (C3) in-stripe ordering and timing, (C4) a single drift-rate solution with a plausible beam-response envelope, and (C5) cross-pass consistency where a position is illuminated in both passes.

After (C1) gating, the vast majority of raw \texttt{turboSETI} hits are removed by (C2), reflecting array-illuminating or band-occupancy RFI. The remaining gated candidates are then eliminated primarily by (C3) and (C4), which require the deterministic in-stripe schedule and a coherent $\dot{\nu}$ solution within gOTWs. Where geometry provides illumination in both passes, (C5) rejects the few residual events lacking consistent drift or timing across passes. No event satisfies the full set of criteria, yielding a robust null detection under complementary two-pass coverage.

\begin{figure*}[t]
\centering

\begin{minipage}[t]{0.485\linewidth}
  \centering
  \includegraphics[width=\linewidth]{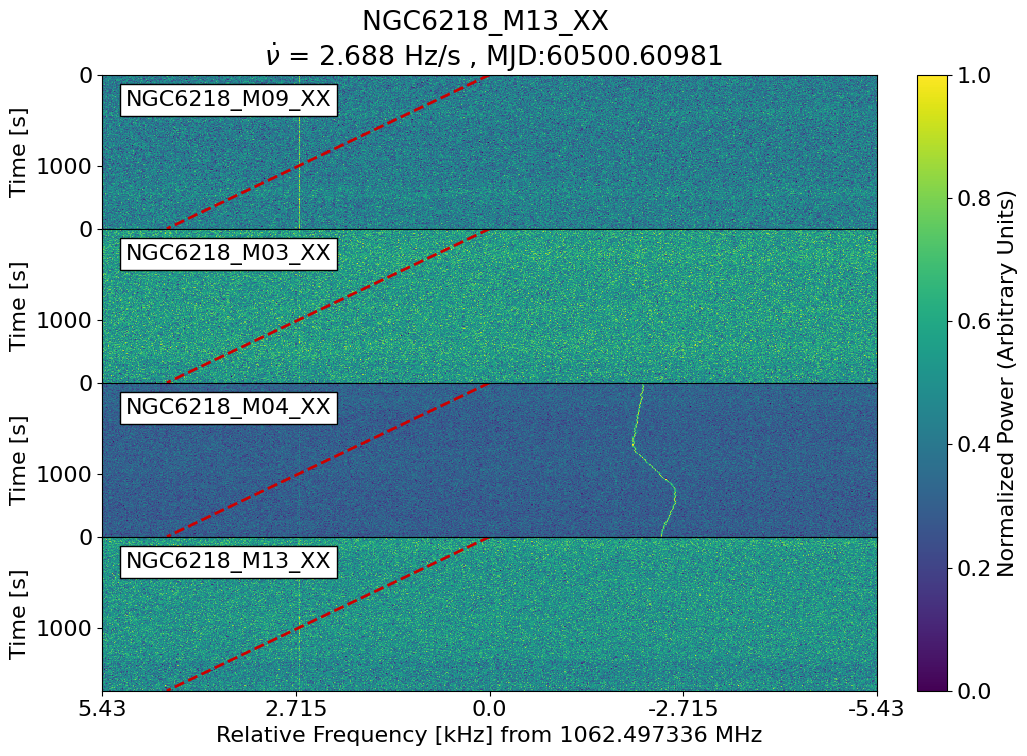}
  \vspace{2pt}\\
  \textbf{(a)} NGC\,6218 (M12), M13, XX. $\dot{\nu}=+2.688$ Hz s$^{-1}$ at $f_0\simeq1062.497$ MHz. 
  High–drift narrowband feature brightens across multiple beams in the same window → violates C1 (gOTW) and C2 (simultaneity).
\end{minipage}\hfill
\begin{minipage}[t]{0.485\linewidth}
  \centering
  \includegraphics[width=\linewidth]{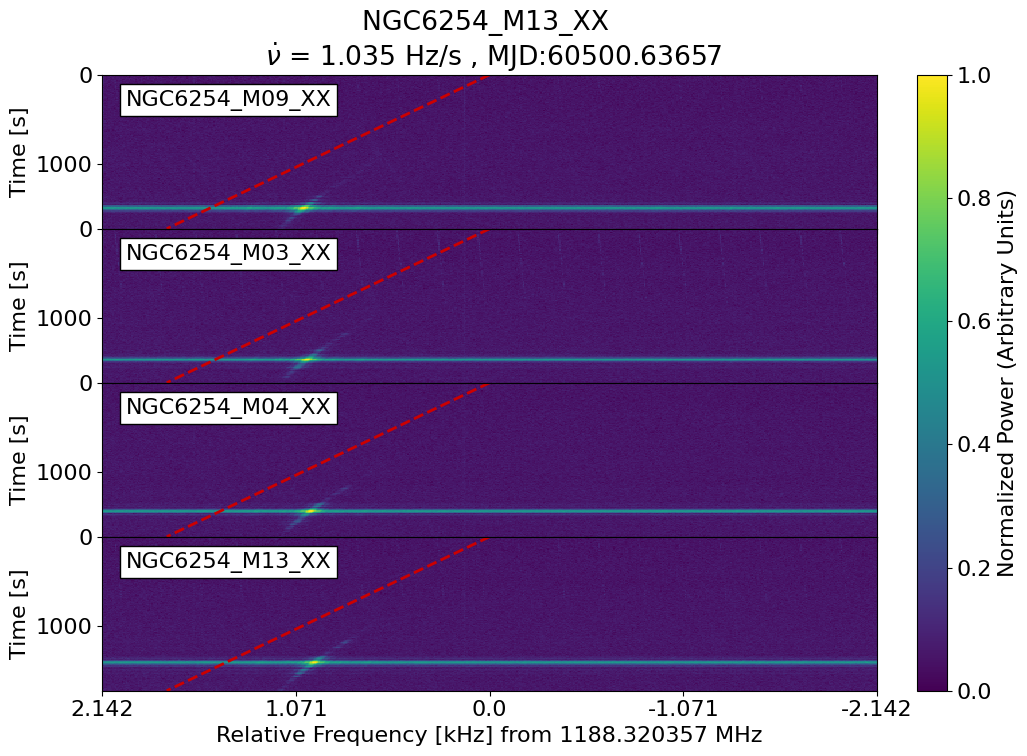}
  \vspace{2pt}\\
  \textbf{(b)} NGC\,6254 (M10), M13, XX. $\dot{\nu}=+1.035$ Hz s$^{-1}$ at $f_0\simeq1188.32$ MHz. 
  Extremely narrow carrier with identical timing/amplitude across beams → fails C1–C3; backend/telecom-like.
\end{minipage}

\vspace{0.8em}

\begin{minipage}[t]{0.485\linewidth}
  \centering
  \includegraphics[width=\linewidth]{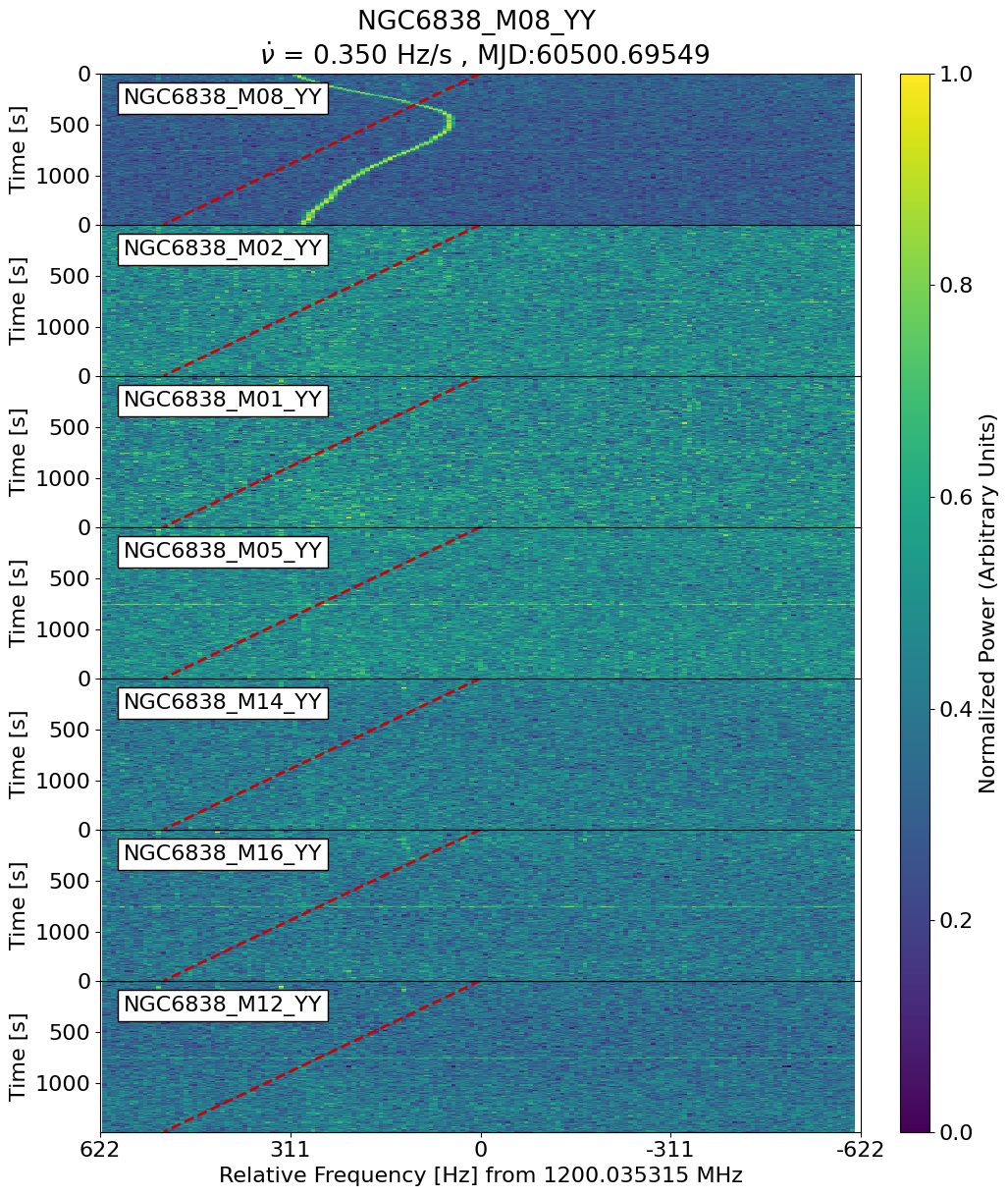}
  \vspace{2pt}\\
  \textbf{(c)} NGC\,6838 (M71), M08, YY. $\dot{\nu}=+0.350$ Hz s$^{-1}$ at $f_0\simeq1200.035$ MHz. 
  Bright tone confined to one feed with a kinked track and no cross-beam recurrence → fails C4; instrumental/near-field.
\end{minipage}\hfill
\begin{minipage}[t]{0.485\linewidth}
  \centering
  \includegraphics[width=\linewidth]{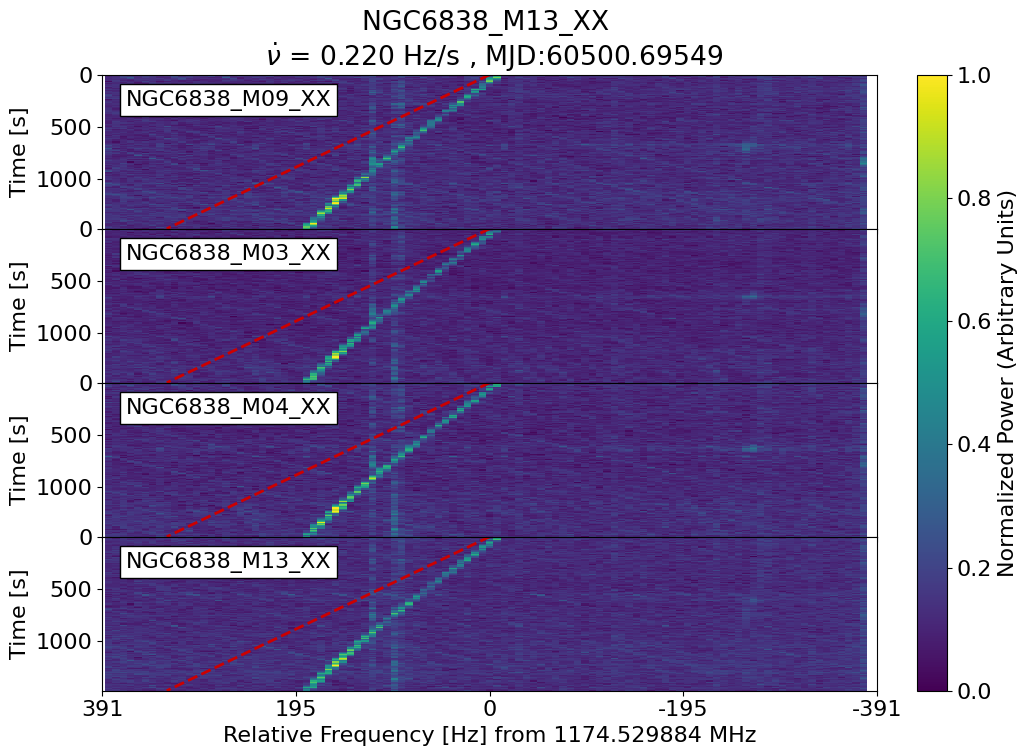}
  \vspace{2pt}\\
  \textbf{(d)} NGC\,6838 (M71), M13, XX. $\dot{\nu}=+0.220$ Hz s$^{-1}$ at $f_0\simeq1174.53$ MHz. 
  Coherent slopes in several beams at the same time/frequency → array-illuminating; violates C2 and C3.
\end{minipage}

\caption{\textbf{Representative RFI morphologies in FAST GC observations.}  
All four examples fail the MBPS verification sequence (C1–C5): (a,b) multibeam simultaneity with flat gain envelopes (C1–C3); (c) single-feed, irregular track (C4); (d) array-wide coherence inconsistent with in-stripe sequencing (C2–C3).}
\label{fig:rfi_panel}
\end{figure*}

\subsection{RFI Environment at FAST}\label{sec:rfi}
The hit density peaks within well-known L-band allocations: aviation SSR/transponder (1030--1140\,MHz) and multiple 
GNSS carriers (e.g., 1176, 1207, 1227, 1268, 1381\,MHz). We also observe telemetry/radar activity near 1250--1300\,MHz and 
downlink spillover toward 1.38--1.40\,GHz. Representative cases (array-wide GNSS tones, out-of-sequence drifters, short 
bursts, and narrow instrumental lines) are rejected by the MBPS criteria in \S\ref{sec:mbps-verify}.

\begin{table*}[t]
\centering
\caption{EIRP thresholds per globular cluster: single illuminated crossing and non-coherent stacking across illuminated crossings (up to 5 per stripe).}
\label{tab:eirp_limits}
\begin{tabular}{lcccc}
\hline
\hline
Cluster & $d$ (kpc) &
\multicolumn{1}{c}{Single crossing (60 s)} &
\multicolumn{1}{c}{Stacked (up to 5 crossings)} \\
\cline{3-3}\cline{4-4}
 &  & ${\rm EIRP}_{\min}$ & ${\rm EIRP}_{\min,5}$ \\
\hline
NGC\,6838 (M71)      & 4.0 & \textcolor{darkgray}{7.2} & \textcolor{darkgray}{3.2} \\
NGC\,6254 (M10)      & 4.4 & \textcolor{darkgray}{8.7} & \textcolor{darkgray}{3.9} \\
NGC\,6218 (M12)      & 4.8 & \textcolor{darkgray}{10.3} & \textcolor{darkgray}{4.6} \\
IC\,1276 (Pal 7)     & 5.4 & \textcolor{darkgray}{13.1} & \textcolor{darkgray}{5.9} \\
NGC\,6171 (M107)     & 6.4 & \textcolor{darkgray}{18.4} & \textcolor{darkgray}{8.2} \\
\hline
\end{tabular}
\begin{flushleft}
\footnotesize
\textbf{Notes.} Values are in $10^{15}$\,W (\textcolor{darkgray}{total EIRP for unresolved narrowband tones integrated over a single 7.5\,Hz channel}). Distances $d$ from the Harris catalog (2010 edition) as cited in the manuscript. 
Single-crossing thresholds adopt ${\rm SEFD}=1.5$\,Jy, ${\rm S/N}=10$, $N_{\rm pol}=2$, $\Delta\nu\simeq7.5$\,Hz, $t_{\rm int}\simeq60$\,s, yielding $S_{\min}\simeq0.50$\,Jy and \textcolor{darkgray}{${\rm EIRP}_{\min}=4\pi d^2 S_{\min}\Delta\nu$} (see §\ref{sec:sensitivity}). 
Stacked values assume \emph{non-coherent} combination across up to five illuminated crossings within a BG stripe (equal weights; unmasked), i.e., $S_{\min,5}=S_{\min}/\sqrt{5}$; where geometry provides fewer crossings, sensitivity scales as $\sqrt{N_{\rm cross}}$. When both passes illuminate the same sky position, additional crossings can be combined under the same rule (see §\ref{sec:mbps-verify}).
\end{flushleft}
\end{table*}

\section{Discussion}\label{sec:discussion}

\subsection{Technosignature prevalence in metal-rich globular clusters}
Our FAST MBPS search toward five nearby, comparatively metal-rich GCs produced a robust null result: no credible, continuous, narrowband detections survived the spatio–temporal MBPS filters. Using the sensitivity framework established in \S\ref{sec:sensitivity} with \({\rm SEFD}=1.5\)\,Jy, \({\rm S/N}=10\), \(N_{\rm pol}=2\), \(\Delta\nu\simeq 7.5\)\,Hz, and \(t_{\rm OTW}\simeq 60\)\,s, the single-crossing threshold is \(S_{\min}\simeq 0.50\)\,Jy and \textcolor{darkgray}{\({\rm EIRP}_{\min}\approx (0.72\text{--}1.8)\times 10^{16}\)\,W} for \(d=4\text{--}6.5\)\,kpc (Table~\ref{tab:eirp_limits}). In a simple Poisson view, surveying \(\sim\)several\(\times 10^{5}\) stars across these clusters with zero detections implies a very low fraction of always-on, isotropic transmitters above the quoted EIRP limits, broadly consistent with field-star surveys at similar thresholds \citep{Enriquez2017BL692,Price2020BL1327,Gajjar2021GC}. This provides the first modern upper limits of this kind in GC environments and shows that bright, persistent L-band beacons are absent at the times and duty cycles we sampled.

\subsection{MBPS versus prior ON--OFF/MBCM strategies}\label{sec:mbps-vs-prior}

Classical ON--OFF and multibeam coincidence matching (MBCM) are snapshot tests: a candidate is accepted if it vanishes in an OFF pointing (ON--OFF) or is present in a designated on-beam while absent in others at one epoch (MBCM). Both schemes control RFI but rely on stationarity and a single on-beam geometry, which reduces on-source duty cycle and creates failure modes for intermittency, beam miscentering, or extended fields \citep{tao2022sensitive,luan2023multibeam}.

MBPS converts vetting from snapshot heuristics to a deterministic, geometry-driven classification test \citep{Huang2023MBPS}. During controlled slews, the multibeam footprint imposes a known time–beam illumination sequence; a sky-localized signal must appear only within its predicted illumination windows, be absent from non-illuminated beams at the same times, and step through beams in the order and cadence set by the scan. This yields orthogonal, in situ checks (spatial exclusivity, simultaneous references, and temporal ordering) without sacrificing duty cycle.

The modification introduced here extends MBPS from point targets to an entire field by (i) defining generalized on–target windows (gOTWs) from the intersection of each beam track with a buffered cluster mask  and (ii) enforcing the same timing/ordering logic within each declination stripe (BG--A to BG--D) and across two complementary passes (Sec.~\ref{sec:mbps-verify}). Passes that do not illuminate a position are not penalized, while doubly illuminated positions must show consistent drift and timing; where applicable, non-coherent combination across crossings improves sensitivity by $\propto \sqrt{N_{\rm cross}}$ (Sec.~\ref{sec:sensitivity}). In practice, this framework preserves the array-wide simultaneity veto of MBCM, restores ON--OFF–level rigor without inter-pointing assumptions, and adds a predictive temporal model that typical site or comb RFI cannot mimic. Applied to our data, no event satisfies the full set of MBPS criteria over the illuminated geometry (Sec.~\ref{sec:mbps-null}), yielding a robust null under extended-target conditions.

\subsection{What we were sensitive to---and what we were not}
Our search is optimized for Hz-wide, slowly drifting carriers across 1.05--1.45\,GHz, detected within \(\sim 60\)\,s OTWs and, when geometry permits, stacked across \emph{available} OTB crossings (upper bound \(\sqrt{N_{\rm cross}}\) improvement; \S\ref{sec:sensitivity}). We are intrinsically less sensitive to (i) broadband or rapidly hopped emissions without matched filters, (ii) very low duty-cycle beacons whose emission seldom overlaps an OTW, and (iii) signals outside L-Band. The thresholds in Table~\ref{tab:eirp_limits} therefore function as a link-budget guide: at kiloparsec distances, isotropic transmission is energetically extravagant, whereas beamed systems (large dishes/arrays) can meet the thresholds with far lower transmitter power.

\paragraph{Engineering mapping.}
Because \({\rm EIRP}=P_{\rm tx}G_{\rm tx}\), our limits can be read as required \(P_{\rm tx}\) at assumed \(G_{\rm tx}\). For reference, at 1.3\,GHz a 300\,m dish (\(\eta\sim 0.6\)) has \(G_{\rm tx}\sim 10^{7}\) (70\,dBi), while 100\,m dishes yield \(G_{\rm tx}\sim 10^{6}\) (60\,dBi). Thus \textcolor{darkgray}{\({\rm EIRP}_{\min}\sim 10^{16}\)\,W} maps to \(P_{\rm tx}\sim 10^{8}\text{--}10^{10}\)\,W depending on aperture; by comparison, the historic Arecibo S-band radar achieved \(\sim 10^{13}\)\,W EIRP \citep[e.g.,][]{Siemion2013Kepler}. This framing clarifies where beamed vs. isotropic strategies land relative to our limits.

\subsubsection{Linear-polarization statistics and implications for Stokes-parameter filtering}\label{sec:pol}
As summarized in Table~3, turboSETI reported 274{,}765 raw hits across both linear polarizations (XX: 143{,}890; YY: 130{,}875). The totals differ by only $\sim$10\% overall (XX/YY $=1.099$), and the normalized asymmetry
\[
A \;\equiv\; \frac{\mathrm{XX}-\mathrm{YY}}{\mathrm{XX}+\mathrm{YY}}
\]
is $A_{\rm all}=0.047$. Four of the five targets lie close to parity (e.g., NGC~6838: XX/YY $=1.014$; IC~1276: $0.974$; NGC~6171: $0.984$; NGC~6254: $1.081$), while one outlier (NGC~6218) shows stronger XX occupancy (XX/YY $=1.462$, i.e., $\approx$46\% more XX than YY hits).\footnote{Per-target asymmetries: $A=\{0.0072,\,-0.0133,\,-0.0079,\,0.1875,\,0.0390\}$ in the order \{NGC~6838, IC~1276, NGC~6171, NGC~6218, NGC~6254\}.}

\smallskip
\noindent \textit{Interpretation.} The near-parity of XX and YY totals indicates that the site’s L-band RFI is, on average, weakly linearly polarized in the FAST feed basis. Larger imbalances can occur transiently (band-occupancy changes, elevation/azimuth coupling, or a handful of bright lines dominating counts), as seen toward NGC~6218.

\smallskip
\noindent \textit{Operational implications.} While our pipeline already analyzes XX and YY independently (§3.1–§3.2), these statistics motivate lightweight polarization-aware heuristics that preserve sensitivity to sky-localized tones while suppressing anthropogenic lines:
\begin{enumerate}
\item \emph{XX–YY consistency check inside gOTWs.} Require approximate consistency of per-crossing intensities in XX and YY (e.g., $|A| \lesssim 0.2$ or a $\lesssim$3\,dB ratio) for promotion to human vetting; do not require strict equality.
\item \emph{Use polarization as a prior, not a veto, outside gOTWs.} Strong, persistent linear polarization seen simultaneously in many \emph{reference} beams (outside gOTWs) is a useful RFI flag (reinforcing the array-wide simultaneity veto, C2).
\item \emph{Path to full Stokes.} Recording cross-hands would enable $Q/U/V$ estimation, allowing stricter Stokes-space vetoes (e.g., persistent circular polarization outside gOTWs) without penalizing astrophysical narrowband tones, which are not expected to be consistently aligned with the X/Y axes a priori.
\end{enumerate}
In short, the $\sim$10\% XX–YY difference in aggregate supports treating polarization as a secondary discriminator that modestly tightens triage with minimal risk to genuine, sky-localized candidates.

\subsubsection{Detection probability for intermittent emitters}\label{sec:pdet}
During the MBOTF raster, a sky position is illuminated in gOTWs of duration $t_{\rm OTW}\!\approx\!60$\,s, separated by $\sim\!115$\,s along the five-beam line (Fig.~\ref{fig:scan_strategy}). Within each illuminated crossing we analyze spectra with cadence $\Delta t\!\approx\!10.2$\,s . \emph{Therefore, the single-crossing sensitivity quoted in §\ref{sec:sensitivity} (e.g., $S_{\min}\!\approx\!0.50$\,Jy for $t_{\rm OTW}\!\approx\!60$\,s) applies to continuous emission within the window.}
 
\paragraph{Window-level detection model.}
Let $\delta_{60}\in[0,1]$ denote the probability that a randomly placed $t_{\rm OTW}{=}\,60$\,s window \emph{contains any emission} from the transmitter (i.e., the window-level occupancy), and let $N_{\rm win}$ be the number of illuminated windows at the candidate sky position across the available passes. If window occupancies are independent,\footnote{Independence is appropriate when the emitter’s on/off process decorrelates on $\lesssim$minute timescales or when any periodicity is not commensurate with the $\sim\!115$\,s spacing between on-target crossings. For strictly periodic bursters with $P$ near integer multiples of $\sim\!115$\,s, window occupancies can be correlated; see the periodic mapping below.} the probability of at least one detection is
\begin{equation}
P_{\rm det} \;=\; 1 - (1-\delta_{60})^{N_{\rm win}},
\label{eq:pdet}
\end{equation}
\noindent \textit{Representative cases.} Using Eq.~\ref{eq:pdet} with $N_{\rm win}{=}5$: for $\delta_{60}{=}0.50$, $P_{\rm det}{=}0.96875$ (97\%); for $\delta_{60}{=}0.10$, $P_{\rm det}{\approx}0.41$; for $\delta_{60}{=}0.05$, $P_{\rm det}{\approx}0.23$; and for $\delta_{60}{=}0.01$, $P_{\rm det}{\approx}0.049$. With ten windows (e.g., both passes illuminating), the corresponding values are $0.999$, $0.65$, $0.40$, and $0.096$, respectively.

\paragraph{Within-window duty fraction and S/N dilution.}
If emission is present in a window but only for a fraction $f\in[0,1]$ of its duration (bursty or intermittent within the 60\,s), the effective integration is $f\,t_{\rm OTW}$. From the radiometer equation (§3.3), the per-window detection S/N scales as $\sqrt{f}$ relative to continuous emission. Equivalently, the \textcolor{darkgray}{flux density} (and EIRP) threshold for detection increases by $1/\sqrt{f}$:
\begin{equation}
S_{\min}(f) \;=\; \frac{S_{\min}(f{=}1)}{\sqrt{f}} \quad \Longrightarrow \quad \mathrm{EIRP}_{\min}(f) \;=\; \frac{\mathrm{EIRP}_{\min}(f{=}1)}{\sqrt{f}}.
\label{eq:fdil}
\end{equation}
For example, if an emitter is active for $f{=}0.25$ of a window (15\,s out of 60\,s), the required \textcolor{darkgray}{flux density}/EIRP for the same S/N increases by a factor of 2. Note that our $\Delta t\!\approx\!10.2$\,s cadence implies $\sim$6 time samples per window; extremely short bursts confined to one sample ($f{\sim}1/6$) incur a $\sim\!2.45\times$ S/N reduction relative to continuous emission in that window (§2.2).

\paragraph{Periodic emitters: mapping to $\delta_{60}$.}
For a strictly periodic transmitter with period $P$ and on-duration $w$ per cycle (duty $d{=}w/P$), and a random phase relative to the raster timing, the probability that a 60\,s window intersects the on-phase is
\begin{equation}
\delta_{60} \;=\; \min\!\left[1,\,\frac{w+t_{\rm OTW}}{P}\right] \;\; \xrightarrow{\,w\ll t_{\rm OTW}\,}\;\; \min\!\left[1,\,\frac{t_{\rm OTW}}{P}\right].
\label{eq:delta_periodic}
\end{equation}
Illustrations (narrow-burst limit $w\!\ll\!60$\,s, single on-segment per $P$): $P{=}120$\,s $\Rightarrow \delta_{60}{\approx}0.50 \Rightarrow P_{\rm det}{=}0.97$ for $N_{\rm win}{=}5$; $P{=}300$\,s $\Rightarrow \delta_{60}{\approx}0.20 \Rightarrow P_{\rm det}{\approx}0.67$; $P{=}600$\,s $\Rightarrow \delta_{60}{\approx}0.10 \Rightarrow P_{\rm det}{\approx}0.41$. If $P$ approaches an integer multiple of the $\sim\!115$\,s separation between adjacent on-target crossings, $\delta_{60}$ remains as above but occupancies across windows can become correlated, slightly modifying Eq.~\ref{eq:pdet}; multi-epoch observations mitigate such commensurabilities.

\paragraph{Implication.}
Our configuration thus strongly favors continuous or frequently active narrowband transmitters, while low-duty-cycle or ultrashort-burst beacons can evade detection unless they overlap one of the $\sim$60\,s windows with sufficient within-window duty fraction $f$ to exceed the radiometer threshold. Increasing the number of illuminated windows via complementary passes and multi-epoch repeats raises $N_{\rm win}$ and improves $P_{\rm det}$ accordingly.

\subsection{Figure of Merit and placement on the CW rate--sensitivity plane}

Following the continuous--wave (CW) comparison framework popularized by \citet{Enriquez2017BL692}, we assess our survey in the plane of sensitivity versus coverage: the abscissa is $\log_{10}(EIRP_{\min}/{\rm W})$ and the ordinate is the \emph{transmitter rate}, defined as the per--star, per--fractional--bandwidth upper limit $(N_\star \,\nu_{\rm rel})^{-1}$. In this space, moving left indicates sensitivity to weaker transmitters (smaller $EIRP_{\min}$) while moving down corresponds to tighter occurrence limits (larger $N_\star \nu_{\rm rel}$).

\begin{figure}[t]
\centering
\includegraphics[width=0.78\linewidth]{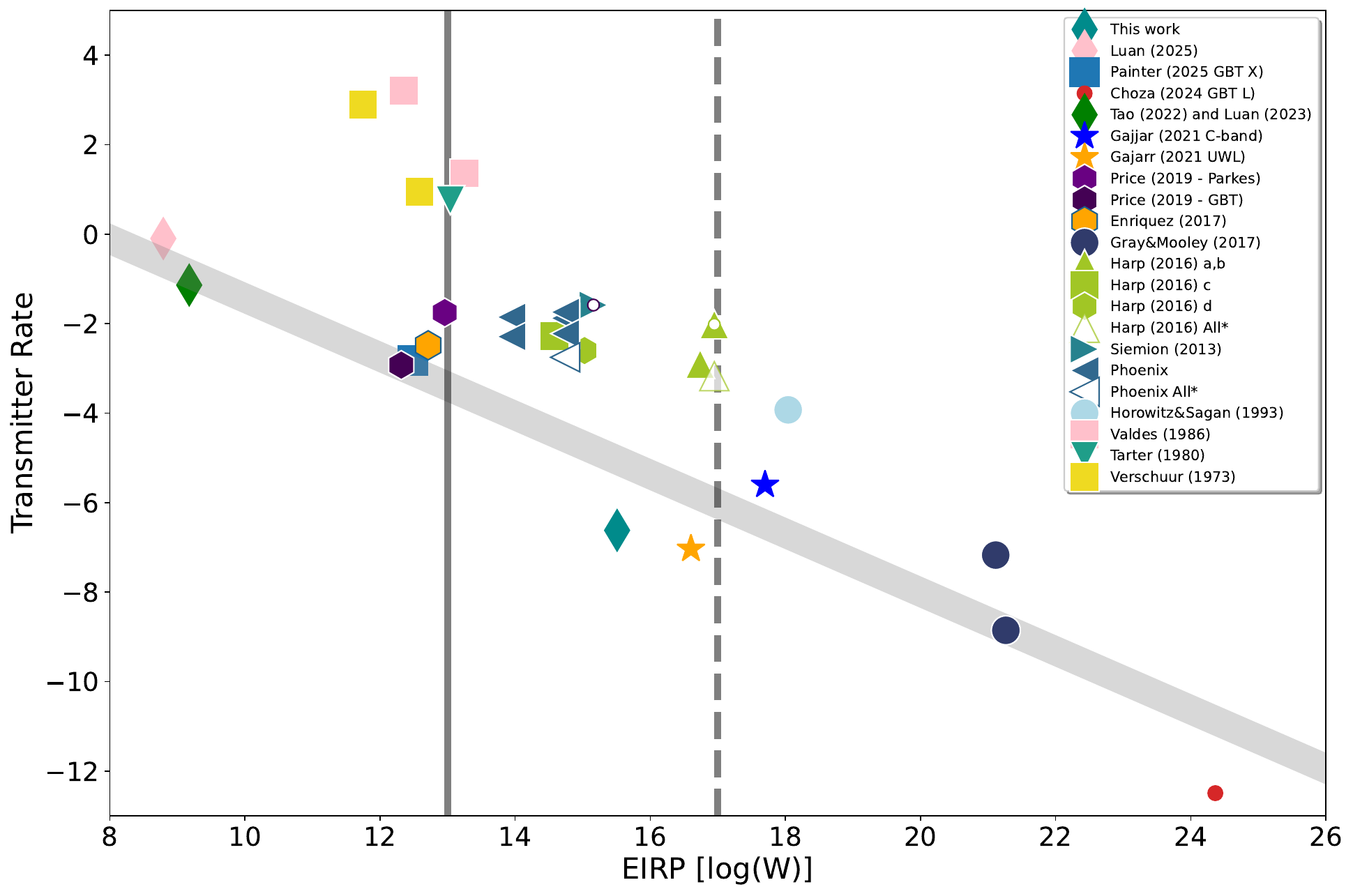}
\caption{Comparison of technosignature survey limits in the plane of transmitter rate versus sensitivity, following the framework of \citet{Enriquez2017BL692}. 
    The abscissa represents the minimum detectable Equivalent Isotropic Radiated Power ($\mathrm{EIRP}_{\min}$), and the ordinate denotes the upper limit on transmitter rate, defined as $(N_{\star} \nu_{\mathrm{rel}})^{-1}$. 
    The \textbf{dark cyan diamond} marks the constraints established by this work (FAST GC Survey Phase I). 
    Historical and contemporaneous limits are plotted for comparison, with data drawn from: 
    \citet{Verschuur1973}, \citet{Tarter1980}, \citet{Valdes1986}, \citet{HorowitzSagan1993}, Project Phoenix \citep{Backus2002}, \citet{Siemion2013Kepler}, \citet{Harp2016}, \citet{GrayMooley2017}, \citet{Price2020BL1327}, \citet{Gajjar2021GC}, \citet{tao2022sensitive}, \citet{luan2023multibeam}, \citet{Choza2024},  \citet{Painter2025} and \citet{Luan_2025}.
    The gray band traces the characteristic sensitivity--coverage trend of these prior searches.
    Our survey's position in the lower-left quadrant highlights its joint optimization of depth (via FAST's high gain) and breadth (via the high stellar density of globular clusters and wide L-band coverage).}
\label{fig:enriquez_plane}
\end{figure}

\paragraph{Where this work lands.}
In Fig.~\ref{fig:enriquez_plane}, the \textbf{dark--cyan diamond} marks our FAST GC survey. It lies \textbf{below and to the left} of the gray trend derived from historical CW searches, indicating a regime that is simultaneously \emph{deep} and \emph{broad}. The leftward placement is set by our L-band thresholds of \textcolor{darkgray}{$EIRP_{\min}\!\approx\!(0.72\text{--}1.8)\times10^{16}$\,W} for single illuminated crossings, with stacking across multiple crossings achieving \textcolor{darkgray}{$\approx(3.2\text{--}8.2)\times10^{15}$\,W} for the nearest targets (Table~4; §3.3). The downward placement reflects the large product $N_\star\nu_{\rm rel}$ delivered by FAST’s wide L-band coverage and the stellar densities of the five targeted clusters, which together drive the per-star, per-bandwidth rate to low values. 

\paragraph{Figure of merit.}
A convenient scalar summary is the CW transmitter figure of merit (CWTFM; \citealp{Enriquez2017BL692}),
\[
{\rm CWTFM}\;\propto\;\frac{EIRP_{\min}}{N_\star\,\nu_{\rm rel}} ,
\]
normalized in the literature so that an Arecibo-benchmark survey has ${\rm CWTFM}=1$. Using our survey totals and L-band coverage, our Arecibo-normalized CWTFM is \textbf{$\ll 1$}, i.e., among the most complete CW searches to date when judged by combined sensitivity and sky-population coverage. 

\paragraph{Why we are lower--left.}
Three ingredients move this survey into the lower-left corner of the Enriquez plane:
(i) \textit{Sensitivity}: FAST’s SEFD $\simeq1.5$\,Jy, $\sim\!60$\,s on-target windows per crossing, and non-coherent stacking across multiple illuminated crossings reduce $EIRP_{\min}$ into the \textcolor{darkgray}{$10^{15.8}$–$10^{16.3}$\,W} range. 
(ii) \textit{Coverage}: multibeam on-the-fly rastering through dense globular-cluster fields yields a large $N_\star$ within the same observing time budget. 
(iii) \textit{Bandwidth}: wide L-band fractional coverage increases $\nu_{\rm rel}$ even after conservative masking of saturated allocations, further tightening the rate axis. 

\paragraph{Implication.}
Being left–and–below the historical gray band means our null result excludes \emph{persistent, isotropic} L-band beacons above \textcolor{darkgray}{$\sim\!10^{16}$\,W} EIRP in the observed epochs for \emph{a large number of stars} in each cluster. For beamed systems, the corresponding transmitter powers are far lower and can be read directly from the link-budget mapping in §5.3. 

\subsection{Conclusions}\label{sec:conclusions}

We have conducted a pilot technosignature search toward five Milky Way globular clusters with FAST using an MBOTF raster and an MBPS analysis adapted to extended targets via gOTW gating and per–BG, per–pass geometric verification. After array-wide simultaneity vetoes and in-stripe timing checks across the two complementary passes, no event satisfies the full MBPS geometry, yielding a robust null detection. Within the survey configuration, the single-crossing sensitivity is $S_{\min}\!\simeq\!0.50$\,Jy, corresponding to \textcolor{darkgray}{${\rm EIRP}_{\min}\!\approx\!(0.72$–$1.8)\!\times\!10^{16}$\,W} for $d=4$–$6.5$\,kpc, with non-coherent stacking providing the expected $\sqrt{N_{\rm cross}}$ gain where geometry allows (Table~\ref{tab:eirp_limits}). 

These results disfavor bright, persistent, isotropic L-band beacons above the quoted thresholds during our epochs, and they delimit the time–frequency–geometry volume within which future searches should expand. Interpreted conservatively, the outcome reflects parameter-space incompleteness (duty cycle, drift range, morphology, or band); interpreted astrophysically, it is consistent with a scarcity of high-EIRP transmitters in the ancient, dynamically dense, and typically metal-poor environments of globular clusters. 

Operationally, this pilot establishes a reproducible FAST blueprint for extended-field SETI—BG/gOTW gating, deterministic geometric validation, and scalable stacking—that now underpins the staged program outlined in §5.4. Executing the next phases (first $\sim$half, then all $\sim$45 FAST-accessible clusters) will convert the present demonstration into a statistically complete census, yielding either the first positive evidence or the most stringent upper limits to date on narrowband transmitters in these systems.

\section{Acknowledgment}

We appreciate the highly positive comments and constructive suggestions from the referee, which greatly improve the paper. This work was supported by the National Key R$\&$D Program of China, No.2024YFA1611804 and the China Manned Space Program with grant No. CMS-CSST2025-A01 and Shandong Provincial Natural Science Foundation (ZR2024QA180) and Scientific Research Fund of Dezhou University (4022504019).This work made use of the data from FAST (Five-hundred-meter Aperture Spherical radio Telescope). FAST is a Chinese national mega-science facility, operated by National Astronomical Observatories, Chinese Academy of Sciences. In the end, the authors thank Shou-Zhi Wang of Beijing Normal University for his guidance on globular clusters.

\bibliography{sample701}{}

@article{DiStefano2016,
  author       = {Di Stefano, Rosanne and Ray, Askar},
  title        = {Globular Clusters as Cradles of Life and Advanced Civilizations},
  journal      = {The Astrophysical Journal},
  year         = {2016},
  volume       = {827},
  number       = {1},
  pages        = {54},
  doi          = {10.3847/0004-637X/827/1/54},
}

@article{FischerValenti2005,
  author       = {Fischer, Debra A. and Valenti, Jeff},
  title        = {The Planet–Metallicity Correlation},
  journal      = {The Astrophysical Journal},
  year         = {2005},
  volume       = {622},
  number       = {2},
  pages        = {1102--1117},
  doi          = {10.1086/428383},
}

@article{Sigurdsson2003,
  author       = {Sigurdsson, Steinn and Richer, Harvey B. and Hansen, Brad M. S. and Stairs, Ingrid H. and Thorsett, Stephen E.},
  title        = {A Young White Dwarf Companion to Pulsar B1620–26: Evidence for Early Planet Formation},
  journal      = {Science},
  year         = {2003},
  volume       = {301},
  number       = {5630},
  pages        = {193--196},
  doi          = {10.1126/science.1086326},
}

@article{BeerKingPringle2004,
  author       = {Beer, Matthias E. and King, Andrew R. and Pringle, J. E.},
  title        = {Planet formation in globular clusters?},
  journal      = {Monthly Notices of the Royal Astronomical Society},
  year         = {2004},
  volume       = {355},
  number       = {1},
  pages        = {1244--1248},
  doi          = {10.1111/j.1365-2966.2004.08399.x},
}

@article{Nan2011,
  author       = {Nan, Rendong and Li, Di and Jin, Chengjin and Wang, Qiming and Zhu, Liang and Zhu, Wenbai and Zhang, Hao and Yue, Youling and Qian, Lei},
  title        = {The Five-hundred-meter Aperture Spherical radio Telescope (FAST) Project},
  journal      = {International Journal of Modern Physics D},
  year         = {2011},
  volume       = {20},
  number       = {6},
  pages        = {989--1024},
  doi          = {10.1142/S0218271811019335},
}

@article{Jiang2020,
  author       = {Jiang, Peng and Yue, Youling and Gan, Hui and Yao, Rongbing and Li, Hongfei and Pan, Zhichen and Sun, Jinghai and Yu, Di and Liu, Xiang and Tang, Nanyi and Qian, Lei and Lu, Jiyuan and Yan, Jingjun and Peng, Bo and Nan, Rendong and Li, Di},
  title        = {Commissioning Progress of the FAST},
  journal      = {Research in Astronomy and Astrophysics},
  year         = {2020},
  volume       = {20},
  number       = {5},
  pages        = {064},
  doi          = {10.1088/1674-4527/20/5/64},
}

@misc{Li2016,
  author       = {Li, Di},
  title        = {FAST in Space: Considerations for a Multibeam SETI Survey},
  year         = {2016},
  eprint       = {1612.09372},
  archivePrefix= {arXiv},
  primaryClass = {astro-ph.IM},
  url          = {https://arxiv.org/abs/1612.09372},
}

@article{Jiang2020FAST,
  author       = {Jiang, P. and Tang, N.-Y. and Hou, L.-G. and Liu, M.-T. and Krco, M. and Qian, L. and Sun, J.-H. and Ching, T.-C. and Liu, B. and Duan, Y. and Yue, Y.-L. and Gan, H.-Q. and Yao, R. and Li, H. and Pan, G.-F. and Yu, D.-J. and Liu, H.-F. and Li, D. and Peng, B. and Yan, J. and FAST Collaboration},
  title        = {The fundamental performance of {FAST} with 19-beam receiver at {L} band},
  journal      = {Research in Astronomy and Astrophysics},
  year         = {2020},
  volume       = {20},
  number       = {5},
  pages        = {064},
  doi          = {10.1088/1674-4527/20/5/64},
  eprint       = {2002.01786},
  archivePrefix= {arXiv}
}

@article{Siemion2013Kepler,
  author       = {Siemion, A. P. V. and Demorest, P. and Korpela, E. and others},
  title        = {A 1.1--1.9 {GHz} {SETI} Survey of the Kepler Field. I. A Search for Narrow-band Emission from Select Targets},
  journal      = {The Astrophysical Journal},
  year         = {2013},
  volume       = {767},
  number       = {1},
  pages        = {94},
  doi          = {10.1088/0004-637X/767/1/94},
  eprint       = {1302.0845},
  archivePrefix= {arXiv}
}

@article{Enriquez2017BL692,
  author       = {Enriquez, J. E. and Siemion, A. P. V. and Foster, G. and others},
  title        = {The Breakthrough Listen Search for Intelligent Life: 1.1--1.9 {GHz} Observations of 692 Nearby Stars},
  journal      = {The Astrophysical Journal},
  year         = {2017},
  volume       = {849},
  number       = {2},
  pages        = {104},
  doi          = {10.3847/1538-4357/aa8d1b},
  eprint       = {1709.03491},
  archivePrefix= {arXiv}
}

@article{Price2020BL1327,
  author       = {Price, D. C. and Enriquez, J. E. and Brzycki, B. and others},
  title        = {The Breakthrough Listen Search for Intelligent Life: Observations of 1327 Nearby Stars Over 1.10--3.45 {GHz}},
  journal      = {The Astronomical Journal},
  year         = {2020},
  volume       = {159},
  number       = {3},
  pages        = {86},
  doi          = {10.3847/1538-3881/ab65f1},
  eprint       = {1906.07750},
  archivePrefix= {arXiv}
}

@article{Gajjar2021GC,
  author       = {Gajjar, V. and Perez, K. I. and Siemion, A. P. V. and others},
  title        = {The Breakthrough Listen Search for Intelligent Life Near the Galactic Center. I.},
  journal      = {The Astronomical Journal},
  year         = {2021},
  volume       = {162},
  number       = {1},
  pages        = {33},
  doi          = {10.3847/1538-3881/abfd36},
  eprint       = {2104.14148},
  archivePrefix= {arXiv}
}

@article{Sheikh2020ETZ,
  author       = {Sheikh, S. Z. and Siemion, A. and Enriquez, J. E. and Price, D. C. and Isaacson, H. and Lebofsky, M. and Kalas, P.},
  title        = {The Breakthrough Listen Search for Intelligent Life: A 3.95--8.00 {GHz} Search for Radio Technosignatures in the Restricted Earth Transit Zone},
  journal      = {The Astronomical Journal},
  year         = {2020},
  volume       = {160},
  number       = {1},
  pages        = {29},
  doi          = {10.3847/1538-3881/ab9361},
  eprint       = {2002.06162},
  archivePrefix= {arXiv}
}

@article{Huang2023MBPS,
  title={A solution to continuous RFI in narrowband radio SETI with FAST: The MultiBeam Point-source Scanning strategy},
  author={Huang, Bo-Lun and Tao, Zhen-Zhao and Zhang, Tong-Jie},
  journal={The Astronomical Journal},
  volume={166},
  number={6},
  pages={245},
  doi={10.3847/1538-3881/ad06b1},
  year={2023},
  publisher={IOP Publishing}
}

@article{harris1996,
  author       = {Harris, William E.},
  title        = {A Catalog of Parameters for Globular Clusters in the Milky Way},
  journal      = {The Astronomical Journal},
  year         = {1996},
  volume       = {112},
  pages        = {1487},
  note         = {2010 edition, online update at \url{http://physwww.mcmaster.ca/\~harris/mwgc.dat}}
}

@article{tao2022sensitive,
  title={Sensitive multibeam targeted SETI observations toward 33 exoplanet systems with FAST},
  author={Tao, Zhen-Zhao and Zhao, Hai-Chen and Zhang, Tong-Jie and Gajjar, Vishal and Zhu, Yan and Yue, You-Ling and Zhang, Hai-Yan and Liu, Wen-Fei and Li, Shi-Yu and Zhang, Jian-Chen and others},
  journal={The Astronomical Journal},
  volume={164},
  number={4},
  pages={160},
  year={2022},
  publisher={IOP Publishing}
}

@article{luan2023multibeam,
  title={Multibeam Blind Search of Targeted SETI Observations toward 33 Exoplanet Systems with FAST},
  author={Luan, Xiao-Hang and Tao, Zhen-Zhao and Zhao, Hai-Chen and Huang, Bo-Lun and Li, Shi-Yu and Liu, Cong and Wang, Hong-Feng and Liu, Wen-Fei and Zhang, Tong-Jie and Gajjar, Vishal and others},
  journal={The Astronomical Journal},
  volume={165},
  number={3},
  pages={132},
  year={2023},
  publisher={IOP Publishing}
}

@ARTICLE{1989Sci,
       author = {{Dreyer}, J.~L.~E. and {Sinnott}, R.~W.},
        title = "{Book-Review - NGC2000.0 - The Complete New General Catalogue and Index Catalogues of Nebulae and Star Clusters}",
      journal = {Science},
         year = 1989,
        month = jan,
       volume = {246},
        pages = {1066},
       adsurl = {https://ui.adsabs.harvard.edu/abs/1989Sci...246.1066D}
}

@article{Verschuur1973,
    author = {{Verschuur}, G.~L.},
    title = "{A Search for Narrow Band 21-cm Wavelength Signals from Ten Nearby Stars}",
    journal = {Icarus},
    year = 1973,
    volume = {19},
    pages = {329},
    doi = {10.1016/0019-1035(73)90109-7}
}

@article{Tarter1980,
    author = {{Tarter}, J. and {Cuzzi}, J. and {Black}, D. and {Clark}, T.},
    title = "{A High Sensitivity Search for Extraterrestrial Intelligence at 18 cm}",
    journal = {Icarus},
    year = 1980,
    volume = {42},
    number = {1},
    pages = {136--144},
    doi = {10.1016/0019-1035(80)90247-8}
}

@article{Valdes1986,
    author = {{Valdes}, F. and {Freitas}, R.~A.},
    title = "{A Search for Narrowband Signals of Extraterrestrial Origin}",
    journal = {Icarus},
    year = 1986,
    volume = {65},
    number = {2},
    pages = {152--156},
    doi = {10.1016/0019-1035(86)90098-7}
}

@article{HorowitzSagan1993,
    author = {{Horowitz}, P. and {Sagan}, C.},
    title = "{Five Years of Project META: An All-Sky Narrow-Band Radio Search for Extraterrestrial Signals}",
    journal = {\apj},
    year = 1993,
    volume = {415},
    pages = {218},
    doi = {10.1086/173157}
}

@inproceedings{Backus2002,
    author = {{Backus}, P.~R. and {Tarter}, J.~C. and {Dreher}, J.~W. and {Harp}, G.~R. and {Ackermann}, R. and {Davis}, M.~M.},
    title = "{Project Phoenix: High Resolution Search for Extraterrestrial Intelligence}",
    booktitle = {Bioastronomy 2002: Life Among the Stars},
    series = {IAU Symposium},
    volume = {213},
    year = 2002,
    editor = {{Norris}, R. and {Stootman}, F.},
    pages = {343}
}

@article{Harp2016,
    author = {{Harp}, G.~R. and {Richards}, J. and {Shostak}, S. and {Tarter}, J.~C. and {Vakoch}, J. and {Scargle}, J.},
    title = "{Radio SETI Observations of the Anomalous Star KIC 8462852}",
    journal = {\apj},
    year = 2016,
    volume = {825},
    number = {2},
    eid = {155},
    pages = {155},
    doi = {10.3847/0004-637X/825/2/155}
}

@article{GrayMooley2017,
    author = {{Gray}, R.~H. and {Mooley}, K.},
    title = "{A Search for Extraterrestrial Intelligence (SETI) toward the Galactic Anticenter with the Murchison Widefield Array}",
    journal = {\aj},
    year = 2017,
    volume = {153},
    number = {3},
    eid = {110},
    pages = {110},
    doi = {10.3847/1538-3881/153/3/110}
}

@article{Choza2024,
    author = {{Choza}, C. and {Brzycki}, B. and {Siemion}, A.~P.~V. and {Croft}, S. and {Czech}, D. and {DeBoer}, D. and {DeRegt}, S. and {Gajjar}, V. and {Gizani}, N. and {Isaacson}, H. and {Lacki}, B. and {Lebofsky}, M. and {MacMahon}, D.~H.~E. and {Price}, D.~C. and {Sheikh}, S.~Z. and {Webb}, L.},
    title = "{The Breakthrough Listen Search for Intelligent Life: Technosignature Search of 97 Nearby Galaxies}",
    journal = {\aj},
    year = 2024,
    volume = {167},
    number = {1},
    eid = {10},
    pages = {10},
    doi = {10.3847/1538-3881/ad09b4}
}

@article{Painter2025,
    author = {{Painter}, C. and {Croft}, S. and {Lebofsky}, M. and {Andersson}, A. and {Choza}, C. and {Gajjar}, V. and {Price}, D.~C. and {Siemion}, A.~P.~V.},
    title = "{A Novel Technosignature Search in the Breakthrough Listen Green Bank Telescope Archive}",
    journal = {arXiv e-prints},
    year = 2025,
    eid = {arXiv:2412.05786},
    pages = {arXiv:2412.05786},
    doi = {10.48550/arXiv.2412.05786},
    archivePrefix = {arXiv},
    eprint = {2412.05786},
    primaryClass = {astro-ph.IM}
}

@article{Luan_2025,
doi = {10.3847/1538-3881/adbaef},
url = {https://doi.org/10.3847/1538-3881/adbaef},
year = {2025},
month = {mar},
publisher = {The American Astronomical Society},
volume = {169},
number = {4},
pages = {217},
author = {Luan, Xiao-Hang and Huang, Bo-Lun and Tao, Zhen-Zhao and Cui, Yan and Zhang, Tong-Jie and Wang, Pei},
title = {Multibeam SETI Observations Toward Nearby M Dwarfs with FAST},
journal = {The Astronomical Journal},
}
\bibliographystyle{aasjournalv7}

\end{CJK*}

\end{document}